\documentclass[12pt]{article}
\usepackage{verbatim,color,amssymb,bbm,epsfig,amsmath}
\usepackage[section]{placeins}
\usepackage[usenames,dvipsnames]{xcolor}
\usepackage{fancyhdr}
\usepackage[authoryear, sort]{natbib}
\usepackage{multirow}
\usepackage{url}
\usepackage{enumitem}
\usepackage{epsfig}

\allowdisplaybreaks

\newtheorem{theorem}{\underline{\bf Theorem}}
\newtheorem{assumption}{\underline{\bf Assumption}}

\newtheorem{Cor}{\underline{\bf Corollary}}

\newtheorem{remark}{\underline{\bf Remark}}

\def\rR{\mathbb{R}}

\def\boxit#1{\vbox{\hrule\hbox{\vrule\kern6pt  \vbox{\kern6pt#1\kern6pt}\kern6pt\vrule}\hrule}}

\def\log{\hbox{log}}

\def\bse{\begin{eqnarray*}}
\def\ese{\end{eqnarray*}}
\def\be{\begin{eqnarray}}
\def\ee{\end{eqnarray}}
\def\bq{\begin{equation}}
\def\eq{\end{equation}}
\def\bse{\begin{eqnarray*}}
\def\ese{\end{eqnarray*}}
\def\pr{\hbox{pr}}

\def\trans{^{\rm T}}

\def\th{^{th}}
\def\b1e{{\mathbf e}}

\def\trans{^{\rm T}}

\def\th{^{th}}
\def\b1e{{\mathbf e}}

\renewcommand{\hat}{\widehat}
\renewcommand{\tilde}{\widetilde}

\newcommand{\vt}{\vartheta}
\newcommand{\bvt}{\bar{\vt}}
\newcommand{\hvt}{\hat{\vt}}
\newcommand{\tvt}{\tilde{\vt}}
\newcommand{\vp}{\xi}
\newcommand{\hvp}{\hat{\vp}}
\newcommand{\home}{\hat {w}^*}
\newcommand{\hq}{\hat{q}}

\newcommand{\htn}{\hat{\Theta}_n}

\newcommand{\hbe}{\hat{\beta}}

\def\mn{ {\cal N}}
\def\mf{ {\cal F}}
\def\mb{ \mathcal{B}_{n,k}}
\def\mq{ \mathcal{Q}_n}
\def\hmq{ \hat{\mathcal{Q}}_n}
\def\hq{ \hat{\mathcal{Q}}}
\def\xq{X_{\mq}}
\def\xqc{X_{\mq^c}}

\newcommand{\p}{\mathrm{pr}}
\def\sj{\hbox{$\sum_{j=1}^{p_n}$}}
\def\si{\hbox{$\sum_{i=1}^n$}}

\def\sk{\hbox{$\sum_{k=1}^K$}}
\def\skp{\hbox{$\sum_{k=1}^{K'}$}}

\def\th{^{\rm th}}

\def\hve{\hat{\varepsilon}_{i}}

\def\trans{^{\rm T}}

\def\boxit#1{\vbox{\hrule\hbox{\vrule\kern6pt  \vbox{\kern6pt#1\kern6pt}\kern6pt\vrule}\hrule}}

\def\rcom#1 {{\color{red}\bf#1} }
\def\bcom#1 {{\color{blue}\bf#1} }

\def\v{{\varepsilon}}

\def\bse{\begin{eqnarray*}}
\def\ese{\end{eqnarray*}}
\def\be{\begin{eqnarray}}
\def\ee{\end{eqnarray}}

\begin{document}

\thispagestyle{empty}
\baselineskip=28pt

\begin{center}
{\LARGE{\bf Penalized regression with multiple loss functions and selection by vote}}
\end{center}

\baselineskip=12pt

\vskip 2mm
\begin{center}
	Guorong Dai and Ursula U. M\"uller\\
	Department of Statistics, Texas A\&M University, College Station, TX 77843-3143, USA \\
	rondai@stat.tamu.edu and uschi@stat.tamu.edu\\
	\hskip 5mm \\
	\hskip 5mm\\ \hskip 5mm\\
	Raymond J. Carroll\\
	Department of Statistics, Texas A\&M University, College Station, TX 77843-3143, USA \\ and School of Mathematical and Physical Sciences, University of Technology Sydney, Broadway NSW 2007, Australia\\
	carroll@stat.tamu.edu
\end{center}

\begin{center}
{\Large{\bf Abstract}}
\end{center}
This article considers a linear model in a high dimensional data scenario. We propose a process which uses multiple loss functions both to select relevant predictors and to estimate parameters, and study its asymptotic properties. Variable selection is conducted by a procedure called ``vote", which aggregates results from penalized loss functions. Using multiple objective functions separately simplifies algorithms and allows parallel computing, which is convenient and fast. As a special example we consider a quantile regression model, which optimally combines multiple quantile levels. We show that the resulting estimators for the parameter vector are asymptotically efficient. Simulations and a data application confirm the three main advantages of our approach: (a) reducing the false discovery rate of variable selection; (b) improving the quality of parameter estimation; (c) increasing the efficiency of computation.

\baselineskip=12pt

\baselineskip=12pt
\par\vfill\noindent

\par\medskip\noindent
\underline{\bf Short title}: Penalized regression with multiple loss functions

\clearpage\pagebreak\newpage
\pagenumbering{arabic}
\newlength{\gnat}
\setlength{\gnat}{22pt}
\baselineskip=\gnat

\section{Introduction}\label{secintro}
Consider a linear model
\be
Y=X\vt+\varepsilon, \label{homomodel}
\ee
where $Y=(Y_1,\ldots,Y_n)\trans$ is an $n$-dimensional vector of responses, $X=(X_1,\ldots,X_n)\trans$ is an $n\times p$ design matrix of predictors with $X_i=(X_{i1},\ldots,X_{ip})\trans$ for $i=1,\ldots,n$, $\vt=(\vt_1,\ldots,\vt_{p})\trans$ is a $p$-dimensional vector of parameters, and $\varepsilon=(\varepsilon_1,\ldots,\varepsilon_n)\trans$ is an $n$-dimensional vector of independent and identically distributed random errors, which is independent of $X$. We study a scenario with high dimensional data where the dimension of the predictors $p=p_n$ is allowed to be greater than the sample size $n$ and tend to infinity as $n$ increases. We further assume that the model is sparse, i.e.\ only a fraction of the predictors significantly affects the response, while the parameters of the other predictors are zero.

One possible approach to selecting the important predictors and to estimate parameters is penalized regression. Various types of penalties, such as the Lasso \citep{tibshirani1996regression}, the smoothly clipped absolute deviation (SCAD) penalty \citep{fan2001variable}, and the adaptive Lasso \citep{zou2006adaptive}, have been applied to least squares regression. Penalized quantile regression has been considered because of its robustness \citep{wu2009variable,wang2012quantile,fan2014adaptive}. In addition to these articles working with a single loss function, \citet{zou2008composite} introduced a composite quantile regression approach, which combines multiple quantile loss functions. They used an adaptive Lasso penalty to detect sparsity in linear models with a fixed number of parameters. Each of the above loss functions only works well for certain classes of error distributions. To obtain universal optimality, \citet{bradic2011penalized} developed a composite quasi-likelihood function, which approximates the log-likelihood function of the random error by a weighted linear combination of convex loss functions, and adopts a weighted Lasso penalty. The use of multiple loss functions in \citet{zou2008composite} and \citet{bradic2011penalized} improves the efficiency of estimation. However, it does not necessarily reduce the false discovery rate in model selection. In fact, a loss function that yields an efficient estimator for a parameter may not be optimal for variable selection. For example, in Section 5.1 of \citet{bradic2011penalized}, the simulation results show that penalized least squares regression can have higher false discovery rates than methods using other loss functions, even though it attains the Cram\'{e}r-Rao bound because the error distribution is normal. Moreover, solving a penalized combination of multiple loss functions can be computationally complex, requiring complicated algorithms. For high dimensional data, the computation becomes increasingly intensive
as the number of loss functions rises \citep[Section 4.2]{bradic2011penalized}.

To make full use of multiple loss functions and facilitate computation, this article proposes a two-step process for the linear model (\ref{homomodel}): in the first step, the ``vote procedure'', the relevant predictors are determined. Then, in the second step, the parameters of the selected variables are estimated. This is different from the methods above, which generate a sparse estimator in one step by minimizing a penalized objective function. Our method uses multiple loss functions to which a weighted Lasso penalty is added. This yields different sparse (preliminary) estimators for the parameter vector $\vt$. If a component of $\vt$ is identified as nonzero by a sufficient number of models, i.e.\ it has received enough ``votes'', the corresponding predictor is included in the final model. Our approach has smaller variance than most existing methods and excludes unimportant predictors more effectively. It requires no sophisticated algorithms and allows parallel computing to reduce processing time. A further advantage is that one can use different loss functions for variable selection and for parameter estimation, which brings more flexibility.

In the next section we introduce our approach, including the vote procedure, and study its asymptotic properties. Section \ref{secex} focuses on a special example, which optimally combines multiple quantile levels, and proves it is asymptotically efficient when the number of loss functions tends to infinity. Our method is compared with other competing methods by means of simulations in Section \ref{secs}. In Section \ref{secapp} we illustrate our approach with a real data analysis. Section \ref{seccon} concludes the article with a brief summary and a discussion of further questions. All proofs are in the Appendix.

\section{Variable selection and parameter estimation with multiple loss functions}\label{secm}
In the following we will, for convenience of notation, let the lower case letter $c$ represent a generic constant. The lower case letter $r$ denotes a vector of a proper length whose components all equal one and the lower case letter $v$ represents a unit vector of a proper length. For a matrix $B$, let $\lambda_{\max}(B)$ and $\lambda_{\min}(B)$ denote its maximum and minimum eigenvalues. Set $\|B\|=\lambda^{1/2}_{\max}(B\trans B)$, $\|B\|_{\infty}$ equal to the largest absolute value of the entries in $B$ and $\|B\|_{2,\infty}=\sup_{\|v\|=1}\|Bv\|_{\infty}$.

 Assume that in the model (\ref{homomodel}), the parameter vector $\vt$ is sparse, i.e.\ there exists a set $\mq\subset\{1,2,\ldots,p_n\}$ such that $\vt_j\neq 0$ for any $j\in\mq$ and $\vt_j=0$ for any $j\in\mq^c=\{1,2,\ldots,p_n\}\backslash\mq$. Without loss of generality, let $\mq=\{1,2,\ldots,q_n\}$ for some positive sequence $q_n<p_n$. To identify the set $\mq$ we consider $K$ different loss functions $\ell_1(\cdot),\ldots,\ell_K(\cdot)$. A weighted Lasso penalty with weights $d_{k1},\ldots,d_{kp_n}$ and a tuning parameter $\lambda_{n,k}$, is applied to the $k\th$ loss function. Then we obtain preliminary estimators
\be
\tvt_k=(\tvt_{k1},\ldots,\tvt_{kp_n})\trans=\hbox{$\mathop{\arg\min}_{\theta}$}\{\si\ell_k(Y_i-X_i\trans\theta)+n\lambda_{n,k}\sj d_{kj}|\theta_j|\}
\label{ob}
\ee
for $k=1,\ldots,K$, where $\theta=(\theta_1,\ldots,\theta_{p_n})\trans$. For some positive integer $\alpha\leq K$, the set $\mq$ is estimated by the vote procedure, i.e.,
\be
\hmq(\alpha)=\{j\in\{1,2,\ldots,p_n\}:\sk I(\tvt_{kj}\neq 0)\geq\alpha\},
\label{vote}
\ee
where $I(\cdot)$ is the indicator function and $\alpha$ is a threshold.
This means that the $j\th$ component $\vt_j$ of $\vt$ is included in the
model if it receives $\alpha$ or more votes, i.e.\ at least $\alpha$
of the $K$ estimates $\tvt_{1j}, \ldots, \tvt_{Kj}$ are nonzero. In Theorem \ref{selection}, we will see that the choice of $\alpha$ does not affect the result of variable selection asymptotically. When the sample size is finite we recommend cross validation to determine $\alpha$ for simulations and data analysis. A similar vote procedure was employed by \citet{meinshausen2010stability} to develop stability selection  and by \citet{chen2014split} to apply the split and conquer strategy to penalized regression. Unlike these two articles, which calculate multiple estimators on subsets of  data, our method obtains each preliminary estimator (\ref{ob}) on the whole data set. This is desirable when the sample size is is not too large. In addition, the same penalty tuning parameter  is used for all estimators in \citet{meinshausen2010stability}, while we choose a tuning parameter $\lambda_{n,k}$ for each estimator independently.

After variable selection  we use $K'$ different loss functions $\ell'_1(\cdot),\ell'_2(\cdot),\ldots,\ell'_{K'}(\cdot)$ to estimate the parameter vector $\vt$.
For $k=1,\ldots,K'$ we calculate
\be
\hvt_k=(\hvt_{k1},\ldots,\hvt_{kp_n})\trans=\hbox{$\mathop{\arg\min}_{\theta\in\htn(\alpha)}$}\si\ell'_k(Y_i-X_i\trans\theta),
\label{obf}
\ee
where $\htn(\alpha)=\{\theta\in\rR^{p_n}:\theta_j=0 \hbox{ for } j\notin\hmq(\alpha)\}$ is an empirical version of the set $\Theta_n=\{\theta\in\rR^{p_n}:\theta_j=0 \hbox{ for } j\notin\mq\}$. Our final estimator for $\vt$ is
\be
\hvt=\skp\home_k\hvt_k,
\label{fe}
\ee
where $\home=(\home_1,\ldots,\home_{K'})\trans$ is a consistent estimator of the optimal weight vector $w^*=(w^*_1,\ldots,w^*_{K'})\trans$
that minimizes the asymptotic variance; see Theorem \ref{normality}
  and Corollary \ref{oweight} for details. Since we have model selection consistency (see Theorem \ref{selection}),
  we will work with $\hvt_{k\mq}=(\hvt_{k1},\ldots,\hvt_{kq_n})\trans$
  instead of $\hvt_k=(\hvt_{k1},\ldots,\hvt_{kp_n})\trans$ to derive these
  statements.

In the above process multiple loss functions are used not only to increase the efficiency of parameter estimation by the weighted combination (\ref{fe}) but also to improve the result of variable selection by the vote procedure (\ref{vote}). For the calculation of the estimators in (\ref{ob}) and (\ref{obf}), we only need algorithms for the $K$ single objective functions. The multiple minimization procedures can then be conducted in parallel; see Section \ref{secs} for details. The two estimation steps may use different sets of loss functions. For example, if the error distribution is thought to be normal, we can use multiple loss functions for (\ref{ob}) and only the quadratic loss function for (\ref{obf}). For simplicity of notation, in the rest of this section and the next section we assume $K=K'$ and $\ell_k(x)=\ell'_k(x)$. Except for minor differences in notation, the conclusions are exactly the same  if we drop these assumptions.

To study the asymptotic properties of the estimators (\ref{vote}) and (\ref{fe}), we impose the following assumptions:
\begin{assumption}\label{lfd}
  For $k=1,\ldots,K$, let $\psi_k(\cdot)$ be a subdifferential of $\ell_k(\cdot)$ and $\mn_k$ be the set of not differentiable points of $\psi_k(\cdot)$. The distribution of $\varepsilon_1$ satisfies $\p\{\v_1\in(\cup_{1\leq k\leq K}\mn_k)\}=0$.
\end{assumption}

\begin{assumption}\label{lfe}
	For $k=1,\ldots,K$, the function $\psi_k(\cdot)$ satisfies that $E\{\psi_{k}(\varepsilon_1+x)\}=\eta_kx+o(|x|)$ as $|x|\to 0$ for some $\eta_k>0$ and that $E\{|\psi_k(\varepsilon_1)|^m\}\leq c\,m!T^{m-2} $ for any $m\geq 2$ and some constant $T>0$. For $i,j=1,\ldots,K$ and sufficiently small $|x|$, the expectation $E[\{\psi_i(\varepsilon_1+x)-\psi_i(\varepsilon_1)\}\{\psi_j(\varepsilon_1+x)-\psi_j(\varepsilon_1)\}]$ exists and is continuous at $x=0$.
\end{assumption}

\begin{assumption}\label{dim}
	There are constants $\kappa,\nu_0\in(0,1)$ such that $\log\,p_n=O(n^{\kappa})$ and $q_n=O(n^{\nu_0})$.
\end{assumption}

\begin{assumption}\label{dm1}
	Let $\xq=(X_{1\mq},\ldots,X_{n\mq})\trans$ and $\xqc=(X_{1\mq^c},\ldots,X_{n\mq^c})\trans$, where $X_{i\mq}=(X_{i1},\ldots,X_{iq_n})\trans$ and $X_{i\mq^c}=\{X_{i(q_n+1)},\ldots,X_{ip_n}\}\trans$ for $i=1,\ldots,n$. Then, for $k=1,\ldots,K$ and some constant $\nu_1$,
	\bse
	\hbox{$\sup_{\theta\in\mb}$}\|\xq\trans G_k(\theta)\xqc\|_{2,\infty}=O_p(n^{1-\nu_1})\hbox{ and } \hbox{$\sup_{\theta\in\mb}$}\lambda_{\min}^{-1}\{\xq\trans G_k(\theta)\xq\}=O_p(n^{-1}),
	\ese
	where $\mb$ is a $q_n$-dimensional ball centered at $\vartheta_{\mq}=(\vartheta_1,\ldots,\vartheta_{q_n})\trans$ with a radius $\rho_n$ such that $\rho_n^{-1}=o\{n^{(1-\nu_0)/2}\}$, and $G_k(\theta)$ is a $n\times n$ diagonal matrix whose $(i,i)\th$ component is $\partial \psi_k(Y_i-X_{i\mq}\trans\theta+x)/\partial x|_{x=0}$.
\end{assumption}

\begin{assumption}\label{dm2}
	There are constants $0<c_1\leq c_2$ such that $c_1n\leq\lambda_{\min}(\xq\trans \xq)\leq \lambda_{\max}(\xq\trans \xq)\leq c_2n$. In addition, the design matrix satisfies $\|X\|_\infty\leq c$.
\end{assumption}

\begin{assumption}\label{tp}
	For some constant $\nu_2\in [0,1/2)$, the constants $\kappa$, $\nu_0$ and $\nu_1$ satisfy $\kappa<(\nu_0-2\nu_1)_++2\nu_2\leq 1$, and the tuning parameter $\lambda_{n,k}$ satisfies $\lambda_{n,k}^{-1}=o\{n^{1/2-(\nu_0-\nu_1)_+/2-\nu_2}\}$ for $k=1,\ldots,K$.
\end{assumption}

\begin{assumption}\label{pw}
	The weight $d_{kj}$ of the weighted Lasso penalty in $(\ref{ob})$ satisfies $D_{n,k}=\max_{j\in\mq}d_{kj}=o(n^{\nu_1-\nu_0/2})$, $\lambda_{n,k}D_{n,k}=O\{n^{-(1+\nu_0)/2}\}$ and $\lim\inf_{n\to\infty}(\min_{j\in\mq^c}d_{kj})>0$ for $k=1,\ldots,K$.
\end{assumption}

\begin{assumption}\label{sig}
	The nonzero parameters satisfy that $(\hbox{$\min_{j\in\mq}$}|\vt_j|)^{-1}=o\{n^{(1-\nu_0)/2}\}$.
\end{assumption}

Assumptions \ref{lfd} and \ref{lfe} regulate the loss functions in (\ref{ob}) and (\ref{obf}). Common loss functions, such as the square function and the check function, satisfy these conditions. Assumption \ref{dim} is a standard condition on the growth rate of the model size for linear models with a diverging number of parameters, which can also be found in \citet{bradic2011penalized} and \citet{wang2012quantile}, among others. Assumptions \ref{dm1} and \ref{dm2} guarantee good behaviour of the design matrix. Assumptions \ref{tp} and \ref{pw} are imposed on the weighted Lasso penalty to ensure important predictors can be detected. Assumption \ref{sig} excludes situations where the values of nonzero parameters decay too fast. Conditions similar to Assumptions \ref{lfd}-\ref{pw} were required in \citet{bradic2011penalized} for penalized regression with a weighted linear combination of loss functions, while Assumption \ref{sig} is necessary for the nonzero parameters to be identified with probability approaching one by the preliminary estimators (\ref{ob}) and vote procedure (\ref{vote}).

In Theorem \ref{selection}, we first state that, with probability close to one, the preliminary estimator $\tvt_k$ equals
\be
\tvt^o_k=(\tvt^o_{k1},\ldots,\tvt^o_{kp_n})\trans=\hbox{$\mathop{\arg\min}_{\theta\in\Theta_n}$}\{\si\ell_k(Y_i-X_i\trans\theta)+n\lambda_{n,k}\sj d_{kj}|\theta_j|\},
\label{oracular}
\ee
which is the minimizer of the penalized objective function in the set $\Theta_n=\{\theta\in\rR^{p_n}:\theta_j=0\hbox{ for }j\notin\mq\}$, for $k=1,\ldots,K$. This indicates that $\tvt_k$ can exclude the unimportant variables. Then we show that by aggregating multiple such preliminary estimators, i.e.\ $\tvt_1,\ldots,\tvt_K$, the vote procedure (\ref{vote}) owns model selection consistency. This means, with probability tending to one, the procedure can recover the index set of the nonzero parameters $\mq$.
\begin{theorem}\label{selection}
	If Assumptions \ref{lfd}-\ref{sig} are satisfied, then for $k=1,\ldots,K$,
	\bse
	\p\{\tvt_k=\tvt^o_k\}\geq 1-2(p_n-q_n)\exp(-c\,z_n^2),
	\ese
	 with $z_n=n^{(\nu_0-2\nu_1)_+/2+\nu_2}$, where $\nu_0$, $\nu_1$ and $\nu_2$ are the constants in Assumptions \ref{dim}, \ref{dm1} and \ref{tp}. In addition, $\p\{\hmq(\alpha)=\mq\}\to 1$ for any positive integer $\alpha\leq K$.
	
\end{theorem}

Let $\hvt_{k\mq}=(\hvt_{k1},\ldots,\hvt_{kq_n})\trans$ and $\vt_{\mq}=(\vt_{1},\ldots,\vt_{q_n})\trans$. The following theorem gives the asymptotic normality of the nonvanishing part of a weighted estimator $\hvt_{\mq}(w)=\sk w_k\hvt_{k\mq}$ with a general weight vector $w=(w_1,\ldots,w_K)\trans$ satisfying $\sk w_k =1$.
\begin{theorem}\label{normality}
	Let $H$ denote a $K\times K$ matrix whose $(i,j)\th$ entry is $(\eta_i\eta_j)^{-1}E\{\psi_i(\varepsilon_1)\psi_j(\varepsilon_1)\}$ with $\eta_k=\partial E\{\psi_k(\varepsilon_1+x)\}/\partial x|_{x=0}$ being the constant in Assumption \ref{lfe}. Under Assumptions \ref{lfd}-\ref{sig}, we have $v\trans (\xq\trans \xq)^{1/2}\{\hvt_{\mq}(w)-\vt_{\mq}\}\stackrel{d}{\longrightarrow} N(0,w\trans Hw)$, provided the constant $\nu_0$ in Assumption \ref{dim} satisfies $\nu_0 < 1/3$ and $\sup_{\|v_1\|=1,\|v_2\|=1}\si (v_1\trans X_{i\mq} X_{i\mq}\trans v_2)^2=O(n)$.
\end{theorem}

We now specify the optimal weights, which minimizes the asymptotic variance in Theorem \ref{normality}, for the estimator in (\ref{fe}) and show that the limiting distribution is not changed if we replace the optimal weights by a consistent estimator.

\begin{Cor}\label{oweight}
	Suppose the conditions in Theorem \ref{normality} hold. Then the optimal weight vector for the weighted estimator $\hvt_{\mq}(w)$ in Theorem \ref{normality} is $w^*=(r\trans H^{-1}r)^{-1}(H^{-1}r)$. In addition, if $\home$ converges to $w^*$ in probability as $n\to\infty$, the estimator $\hvt_{\mq}(\home)$ converges in distribution as follows: $v\trans (\xq\trans \xq)^{1/2}\{\hvt_{\mq}(\home)-\vt_{\mq}\}\stackrel{d}{\longrightarrow}N\{0,(r\trans H^{-1}r)^{-1}\}$ $(n\to\infty)$.
\end{Cor}

Theorems 1 and 2 establish the model selection consistency and the asymptotic normality of the estimator with arbitrary weights when multiple loss functions are used for both variable selection and parameter estimation. Corollary \ref{oweight} gives the optimal weights and the asymptotic distribution of the estimator (\ref{fe}) when the optimal weights are estimated consistently. The above results justify our approach asymptotically. Its practical advantages in finite samples are illustrated in Sections \ref{secs} and \ref{secapp} with simulations and a data analysis. Although Theorem \ref{selection} indicates that the choice of the threshold $\alpha$ does not affect the result of the vote procedure (\ref{vote}) asymptotically, we recommend cross validation to choose $\alpha$. 
More details are provided in Sections \ref{secs} and \ref{secapp}.

\section{An example}\label{secex}

A special example of the weighted estimator (\ref{fe}) is the estimator from a quantile regression model optimally combining multiple quantile levels, which was considered in Section 3 of \citet{zhao2014efficient} for a fixed number of parameters without sparsity. These authors used the loss function $\ell_k(x)=(x-\beta_k)\{\tau_k-I(x<\beta_k)\}$ for the estimator (\ref{obf}), where $\tau_k=(K+1)^{-1}k$ and $\beta_k$ is the $\tau_k$ quantile of the random error $\varepsilon_1$, which can be estimated along with the slope vector $\vartheta$ as an additional parameter. In this scenario, the $(i,j)\th$ component of the matrix $H$ in Corollary \ref{oweight} is
\be\label{matrixh}
H_{ij}=\{f(\beta_i)f(\beta_j)\}^{-1}\{\min(\tau_i,\tau_j)-\tau_i\tau_j\},
\ee
where $f(\cdot)$ is the density function of the random error $\varepsilon_1$. We show the asymptotic efficiency of $\hvt$ under the following assumption, which regulates the error distribution.
\begin{assumption}\label{density}
	The error density $f(\cdot)$ is positive, twice differentiable and bounded over the set $\mf=\{x:F(x)\in(0,1)\}$, where $F(\cdot)$ is the distribution function of $\v_1$. For some constant $\nu_3\in[0,\infty]$, the function $g(x)=f\{F^{-1}(x)\}$ with $x\in(0,1)$ satisfies
	\bse
	x^{-1}[\{g(x)\}^2+\{g(1-x)\}^2]\to \nu_3 \hbox{ and } x^2\hbox{$\int^{1-x}_x$} \{g''(t)\}^2dt\to 0 \hbox{ as } x\to 0.
	\ese
\end{assumption}
The above assumption is satisfied by most of common continuous distributions. The value of $\nu_3$ is related to the support of the error density. For example, the constant $\nu_3$ equals zero if the support is $(-\infty,\infty)$. When the support has the form $[s_1,s_2]$, $(-\infty,s_2]$ or $[s_1,\infty)$ for some constants $s_1<s_2$, it is easy to see that $\nu_3$ is infinite.

The following theorem gives the limit of the asymptotic variance $(r\trans H^{-1}r)^{-1}$ of the optimally weighted estimator $\hvt_{\mq}(w^*)$ from Corollary \ref{oweight} when the number of quantiles $K$ tends to infinity.
\begin{theorem}\label{efficiency}
	Consider the asymptotic variance in Corollary 1 and suppose Assumption 9 holds true, then the reciprocal of that variance satisfies
	\bse
	\hbox{$\lim_{K\to\infty}(r\trans H^{-1}r)=\int_{\mf}\{f(t)\}^{-1}\{f'(t)\}^2dt$}+\nu_3,
	 \ese
	 where $\nu_3$ is the constant in Assumption \ref{density}.
\end{theorem}
In the above conclusion, the integral on the right-hand side is the Fisher information. Consequently, we see that $(r \trans H^{-1} r)$ coincides with the Fisher information, i.e.\ we have asymptotic efficiency, when the number of quantiles tends to infinity and $\nu_3$ equals zero. This is the case, for example, if the error distribution is normal. For error distributions with $\nu_3$ greater than zero, the limit of $(r \trans H^{-1} r)$ is larger than the Fisher information, i.e.\ the asymptotic variance becomes even smaller. This holds for irregular cases such as the uniform distribution on $[-1,1]$. The above shows that the estimator $\hvt$ is close to being asymptotically efficient if the number of quantiles is large.

\begin{remark}\label{remark1}
	To estimate $H_{ij}$ in (\ref{matrixh}) for the weight vector $\home$, we conduct the following steps:
	\begin{enumerate}
		\item Calculate $\hve=Y_i-X_i\trans\hvt^{(0)}$ for $i=1,\ldots,n$, where $\hvt^{(0)}=K^{-1}\sk\hvt_k$ is a preliminary version of (\ref{fe}).
		
		\item Estimate $f(\beta_k)$ by a kernel estimator $\hat{f}(\hbe_k)=(nh)^{-1}\si \phi\{h^{-1}(\hbe_k-\hve)\}$ for $k=1,\ldots,K$, where $\hbe_k$ is the $\tau_k$ sample quantile of $\hat{\varepsilon}_1,\ldots,\hat{\varepsilon}_n$, $\phi(\cdot)$ is the standard normal density function, and $h=0.9n^{-1/5}\min\{\hbox{SD}(\hat{\varepsilon}_1,\ldots,\hat{\varepsilon}_n),\hbox{IQR}(\hat{\varepsilon}_1,\ldots,\hat{\varepsilon}_n)/1.34\}$ is the rule-of-thumb bandwidth \citep{silverman1986density} with SD and IQR standing for the sample standard deviation and sample interquartile range respectively.
		
		\item Estimate $H_{ij}$ by $\hat{H}_{ij}=\{\hat{f}(\hbe_i)\hat{f}(\hbe_j)\}^{-1}\{\min(\tau_i,\tau_j)-\tau_i\tau_j\}$.
	\end{enumerate}
\end{remark}

\section{Simulations}\label{secs}

In this section we study the numerical performance of our method. We consider samples of size $n=200$ throughout. We draw random vectors $X_1,\ldots, X_n$ independently from a $p$-dimensional multivariate normal distribution with mean zero and a variance-covariance matrix whose $(i,j)\th$ component is $0.5^{|i-j|}$. The full model size is $p=12$ or $p=400$ and the nonzero parameters are $(\vt_1,\vt_2,\vt_5)=(3.0,1.5,2.0)$. The response vector is $Y=X\trans\vt+\varepsilon$. Similar regression models were used for simulations by \citet{zou2008composite} and \citet{bradic2011penalized}, among others. We consider the following error distributions: a t-distribution with two degrees of freedom, T$_2$; a normal distribution, N$(0,3)$; a scale mixture of normals, 0.5N(0,\,6)+0.5N(0,\,6$\times 0.5^6$); a location mixture of normals, 0.5N($-2$,\,1)+0.5N(2,\,1); a gamma distribution, $\Gamma(1,1)$; a double exponential distribution with mean 0 and variance 2; a beta distribution, B$(1,3)$; and a uniform distribution, U$(-3,3)$.

In the objective functions (\ref{ob}) and (\ref{obf}) for the preliminary and the final estimator we set $K=K'=9$ and use the same loss function, namely the check function $\ell_k(x)=\ell'_k(x)=(x-\beta_k)\{\tau_k-I(x<\beta_k)\}$ with $\tau_k=k/10$, $k=1, \ldots, 9$. We call this method ``weighted quantile regression through vote" (WQR-vote). To guarantee the best performance we use an iterative scheme for the preliminary estimator $\tvt_k$. The initial value is $\tvt^{(0)}_{k}=\hbox{$\mathop{\arg\min}_{\theta}$}\{\si\ell_k(Y_i-X_i\trans\theta)+n\lambda_{k}\sj |\theta_j|\}$ and the updates in the $t\th$ iteration are
\be
\tvt_k^{(t)}=\hbox{$\mathop{\arg\min}_{\theta}$}[\si\ell_k(Y_i-X_i\trans\theta)+n\lambda_{k}\sj d_{k}\{\tvt_{kj}^{(t-1)}\}|\theta_j|].
\label{iteration}
\ee
Here $\lambda_{k}d_{k}(x)=\lambda_{k}I(|x|\leq\lambda_{k})+(b-1)^{-1}(b\lambda_{k}-|x|)_+I(|x|>\lambda_{k})
$ is the derivative of the SCAD penalty with $b$ being a constant that is usally set to 3.7 \citep{fan2001variable}. We repeat (\ref{iteration}) until convergence. This process is equivalent to minimizing the objective function with the SCAD penalty \citep{zou2008one}.

\begin{table}
	\centering
	\caption{\baselineskip=12pt Mean numbers of correctly selected nonzero parameters (MNC) and mean numbers of incorrectly selected zero parameters (MNI) of the weighted quantile regression through vote method (WQR-vote), least absolute deviation regression (LADR), least squares regression (LSR) and composite quantile regression (CQR), and the relative efficiency (RE) of the WQR-Vote to the three competing methods for various error distributions. Higher values of  relative efficiency indicate better performance of the WQR-vote in estimation. The full model size is $p=12$;
		The abbreviation SMN stands for a scale mixture of normal distributions 0.5N(0,\,6)+0.5N(0,\,6$\times 0.5^6$); LMN is a location mixture of normal distributions 0.5N($-2$,\,1)+0.5N(2,\,1) and DE denotes the double exponential distribution with mean 0 and variance 2.}
	\vskip 3mm
	\label{table1}
	\begin{tabular}{|llcccccccc|}
		\hline
		& & T$_2$ & N(0,\,3) & SMN & LMN & $\Gamma$(1,\,1) & DE &  B(1,\,3) & U(-3,\,3)   \\*[-.30em]
		WQR-vote & MNC & 3 & 3 & 3 & 3 & 3 & 3  & 3 & 3  \\*[-.60em]
		& MNI & 0 & 0 & 0.03 & 0.01 & 0.04 & 0.01 & 0.03 & 0.02 \\*[-.60em]
		& RE & 1 & 1 & 0.98 & 0.99 & 0.97 & 1  & 1 & 1\\*[-.60em]
		& & & & & & & & & \\*[-.60em]
		LADR & MNC & 3 & 3 & 3 & 3 & 3 & 3  & 3 & 3 \\*[-.60em]
		& MNI & 0.89 & 0.76 & 1.03 & 0.86 & 0.91 & 0.85  & 1.10 & 0.51 \\*[-.60em]
		& RE & 1.01 & 1.42 & 0.95 & 12.55 & 6.15 & 0.89  & 4.70 & 4.36 \\*[-.60em]
		& & & & & & & & & \\*[-.60em]
		LSR & MNC & 2.99 & 3 & 3 & 3 & 3 & 3  & 3 & 3  \\*[-.60em]
		& MNI & 0.86 & 0.46 & 0.65 & 0.57 & 0.51 & 0.42  & 0.57 & 0.44 \\*[-.60em]
		& RE & 5.14 & 0.90 & 6.15 & 2.37 & 6.55 & 1.41  & 2.57 & 1.52 \\*[-.60em]
		& & & & & & & & & \\*[-.60em]
		CQR & MNC & 3 & 3 & 3 & 3 & 3 & 3  & 3 & 3  \\*[-.60em]
		& MNI & 0.74 & 0.86 & 0.78 & 1.03 & 0.91 & 0.74 & 0.76 & 0.96\\*[-.60em]
		& RE & 0.99 & 0.97  & 1.64 & 2.15 & 2.85 & 0.95 & 2.22 & 1.80\\
		
		\hline
	\end{tabular}
\end{table}

To solve (\ref{iteration}) we use the quick iterative coordinate descent algorithm \citep{peng2015iterative}. We minimize the multiple objective functions in (\ref{ob}) and (\ref{obf}) in parallel by using the {\tt foreach} function from the {\tt R} package {\tt foreach}. The tuning parameters in the penalty term and the threshold $\alpha$ for the vote procedure (\ref{vote}) are determined by minimizing prediction errors that are calculated over an independent validation set $\{(\tilde{X}_i,\tilde{Y}_i):i=1,\ldots,n'\}$ with $n'=2,000$, which is generated from the distribution of $(X_1,Y_1)$. We choose the tuning parameter $\lambda_k$ in (\ref{iteration}) by setting $\lambda_k=\arg\min_{a\in\mathcal{L}}\hbox{$\sum_{i=1}^{n'}\ell_k(\tilde{Y}_i-\tilde{X}_i\trans\tvt_{a,k})$}$, where $\mathcal{L}$ is a fine grid and $\tvt_{a,k}$ is the outcome of the iterative process (\ref{iteration}) with $\lambda_k=a$. The candidate set for $\alpha$ is $\mathcal{A}=\{\lceil K/2\rceil,\lceil K/2\rceil+1,\ldots,K-1\}$ and the criterion is $\alpha=\hbox{$\mathop{\arg\min}_{a\in\mathcal{A}}$}\sk\hvp_k\hbox{$\sum_{i=1}^{n'}$} \ell_k(\tilde{Y}_i-\tilde{X}_i\trans\hvt_{a,k})$. Here the parameter estimator is $\hvt_{a,k}=\hbox{$\mathop{\arg\min}_{\theta\in\htn(a)}$}\si\ell_k(Y_i-X_i\trans\theta)$ with $\htn(a)=\{\theta\in\rR^{p_n}:\theta_j=0 \hbox{ for } j\notin\hmq(a)\}$. The vector $\hvp=(\hvp_1,\ldots,\hvp_K)\trans$ is a consistent estimator of the optimal weights $\vp=R^{-1}\psi$, where $R$ is a $K\times K$ matrix with $(i,j)\th$ entry $\{\min(\tau_i,\tau_j)-\tau_i\tau_j\}$ and $\psi=\{f(\beta_1),\ldots,f(\beta_K)\}\trans$ \citep[Section 4.2]{bradic2011penalized}. Recall that $\beta_k$ is the $\tau_k$ quantile of the error distribution for $k=1,\ldots,K$. After obtaining the set $\hq(\alpha)$, we construct the estimator $\hvt$ in (\ref{fe}) as described in Remark \ref{remark1} in Section \ref{secex}. Remark \ref{remark1} also explains how to estimate $f(\beta_k)$ for $k=1,\ldots,K$.

For comparison we consider least absolute deviation regression, least squares regression and composite quantile regression with the same nine quantile levels $1/10,\ldots,9/10$. The iterative penalty in (\ref{iteration}) is also applied to the three competing methods. For each of the four estimators, three indices from 200 simulated data sets are recorded in the following tables:
\begin{enumerate}
\item mean number of correctly selected nonzero parameters;

\item mean number of incorrectly selected zero parameters;


\item relative efficiency $E\{\|\hvt_{\mathrm{oracle}}-\vt\|^2\}/E\{\|\hvt_{\mathrm{WQR-vote}}-\vt\|^2\}$ of the WQR-vote procedure, where $\hvt_{\mathrm{oracle}}$ is the oracle version of the four respective estimators and $\hvt_{\mathrm{WQR-vote}}$ is the estimator from the WQR-vote. Here ``oracle'' means knowing the index set $\mathcal{Q}=\{1,2,5\}$ of the nonzero parameters before estimating and applying no penalty.
\end{enumerate}

\begin{table}
	\centering
	\caption{\baselineskip=12pt
		The table entries are mean numbers of correctly selected nonzero parameters, mean numbers of incorrectly selected zero parameters and the relative efficiency as in Table \ref{table1}. In contrast to Table \ref{table1} we now consider a high dimensional data scenario with the full model size $p=400$.
	}
	\vskip 3mm
	\label{table2}
	\begin{tabular}{|llcccccccc|}
		\hline
		& & T$_2$ & N(0,\,3) & SMN & LMN & $\Gamma$(1,\,1) & DE &  B(1,\,3) & U(-3,\,3)   \\*[-.30em]
		WQR-vote & MNC & 3 & 3 & 3 & 3 & 3 & 3  & 3 & 3  \\*[-.60em]
		& MNI & 0 & 0 & 0 & 0 & 0.01 & 0 & 0 & 0 \\*[-.60em]
		& RE & 1 & 1 & 1 & 1 & 0.99 & 1  & 1 & 1\\*[-.60em]
		& & & & & & & & & \\*[-.60em]
		LADR & MNI & 3 & 3 & 3 & 3 & 3 & 3  & 3 & 3 \\*[-.60em]
		& MNI & 2.41 & 1.53 & 2.59 & 1.06 & 2.74 & 1.70  & 2.81 & 1.14 \\*[-.60em]
		& RE & 1.04 & 1.34 & 1.01 & 13.36 & 5.69 & 0.82  & 5.08 & 4.60 \\*[-.60em]
		& & & & & & & & & \\*[-.60em]
		LSR & MNI & 2.94 & 3 & 3 & 3 & 3 & 3  & 3 & 3  \\*[-.60em]
		& MNI & 7.60 & 1.62 & 1.85 & 4.86 & 0.75 & 1.15  & 0.76 & 1.29 \\*[-.60em]
		& RE & 4.97 & 0.88 & 5.89 & 2.51 & 6.36 & 1.33 & 2.62 & 1.67 \\*[-.60em]
		& & & & & & & & & \\*[-.60em]
		CQR & MNC & 3 & 3 & 3 & 3 & 3 & 3  & 3 & 3  \\*[-.60em]
		& MNI & 7.14 & 7.37 & 6.44 & 7.63 & 6.20 & 6.61 & 5.07 & 7.01\\*[-.60em]
		& RE & 0.97 & 0.95  & 1.71 & 2.27 & 2.72 & 0.94 & 2.20 & 1.93\\
		\hline
	\end{tabular}
\end{table}

In Table \ref{table1} we consider the case with $p=12$ predictors. While all the methods successfully select the three nonzero parameters, the WQR-vote selects far fewer zero parameters incorrectly than the others. Moreover, all the relative efficiency values in the table are either close to one (between 0.89 and 1.01) or clearly larger than one (between 1.41 and 12.55),  which shows the efficiency of the WQR-vote are similar to or much better than those of the oracle versions of the competing methods. With respect to computational speed, on a 2.4 GHz processor, the average processing time of the WQR-vote is 1.35 seconds over the 200 replications when the error distribution is N$(0,3)$, while that of composite quantile regression is 1.55 seconds in the same scenario. The computation time in the other cases is similar. This indicates that when WQR-vote and composite quantile regression consider the same number of quantile levels, the WQR-vote can be faster by conducting the computation in parallel. The comparison is probably not entirely fair, because the multiple quantile loss functions in composite quantile regression are unweighted. A weighted version of composite quantile regression would involve estimation and correction of the weights to improve efficiency \citep[Sections 3.4 and 4.2]{bradic2011penalized}, which takes extra time. In Table \ref{table2} we present simulation results as in Table \ref{table1}, but now for the high dimensional scenario with $p=400$ predictors. When dealing with high dimensional data, the WQR-vote still yields the smallest numbers of incorrect selections for all of the cases and the lowest mean squared errors of estimation for most of the cases. In the 200 simulations with normal errors the average time required for the WQR-vote and the composite quantile regression is 27.12 seconds and 106.41 seconds, respectively, when both methods use the nine quantile levels. The advantage of adopting parallel computing in the WQR-vote procedure is obvious.

\section{A data application}\label{secapp}
In this section we analyze a microarray dataset from \citet{scheetz2006regulation}, which is available at \url{https://www.ncbi.nlm.nih.gov/geo/geo2r/?acc=GSE5680}. The dataset contains the gene expression values of 31,042 probes on 120 rats. We are interested in how the expression of TRIM32, which is related to human hereditary diseases of the retina and corresponding to probe 1389163\_at, depends on that of other genes. To exclude genes without sufficient variability, we first remove probes whose maximum among the 120 rats is less than the $25\th$ percentile of all the expression values or whose range among the 120 rats is less than 1. We then sort the remaining 18,984 probes by the absolute values of their correlation coefficients with the response, 1389163\_at. The top 300 probes are used as predictors in the anlaysis.

To identify important predictors by the vote procedure (\ref{vote}), we set $K=9$ and $\ell_k(x)=x\{k/10-I(x<\beta_k)\}$ in the objective function (\ref{ob}), where $\beta_k$ is the $k/10$ quantile of the error distribution and can be estimated as an extra parameter. Then three different estimators are computed based on one loss function, i.e.\ $K'=1$ in (\ref{obf}): the square function, the absolute value function and the composite check function $\ell'_1(x)=\sum_{m=1}^9x\{m/10-I(x<\beta_m)\}$. For comparison, the data are also analyzed by penalized regression using each of the above three loss functions. The iterative penalty (\ref{iteration}) is applied to all methods.

We first use the entire dataset to fit models. Then the dataset is randomly divided into a training set of 80 observations and a validation set of 40 observations. Models are fitted with the training set and prediction errors are calculated by the loss functions on the validation set. Based on five-fold cross validation, we choose the tuning parameters in the penalty and the threshold $\alpha$ in (\ref{vote}) by the criteria stated in Section \ref{secs}. This is repeated 50 times.

\begin{table}
	\centering
	\caption{\baselineskip=12pt Sizes of selected subsets based on all data and on the 50 randomly generated partitions. We consider least absolute deviation regression (LADR), least squares regression (LSR) and composite quantile regression (CQR). The term LADR-Vote denotes the LADR appoach (without penalty) after variable selection through the vote procedure; LSR-Vote and CQR-Vote are defined analogously.
          }
	\vskip 3mm
	\label{table3}
	\begin{tabular}{|lc|ccc|}
          \hline
         & All Data
          & \multicolumn{3}{c|}{Random partition} \\*[-.60em]
		& Model size &  Model size & \multicolumn{2}{c|}{Prediction error} \\*[-.30em]
		& & & LADR-Vote & 2.68 (0.33)  \\*[-.60em]
		Vote & 7 & 7.54 (4.14) & LSR-Vote & 0.29 (0.08)   \\*[-.60em]
		& & & CQR-Vote & 13.02 (2.71)  \\*[-.30em]
		LADR & 16 & 12.86 (6.84) & & 3.51 (0.49)  \\*[-.60em]
		LSR & 10 & 10.96 (4.36) & & 0.34 (0.09)  \\*[-.60em]
		CQR & 14 & 12.75 (10.35) & & 14.60 (2.36)  \\
		\hline
	\end{tabular}
\end{table}

  Table \ref{table3} gives the sizes of the models that are selected by the vote procedure and by the three competing penalized regression approaches. The left panel provides the sizes that are obtained using all data. The first column of the right panel lists the means and, in parentheses, the standard deviations of the model sizes that are obtained using the 50 randomly generated partitions. The top half of the second column provides the prediction errors of the parameter estimates that are obtained in the second step of our method, i.e.\ after model selection by the vote procedure, using least absolute deviation regression, least squares regression and composite quantile regression. In the bottom half are the prediction errors of the estimators using the penalized versions of the three regression approaches. The prediction errors are calculated using different loss functions. Hence they have different scales and are not comparable across models. Comparable are LADR and LADR-Vote, LSR and LSR-Vote, as well as CQR and CQR-Vote.

Among all methods, the vote method generates the smallest models in the situations of both all data and random partition. 
In addition to achieving more sparsity, the submodel selected by the vote method yields a smaller mean prediction error compared with that selected by the penalized regression approach, when the same loss function is used to estimate the parameters. This confirms the superiority of the vote method in variable selection. Furthermore, this analysis illustrates the flexibility of our approach from employing different loss functions for selection and estimation. One may also apply the method optimally combining multiple quantile levels described in Section \ref{secex} to the data. We do not consider it here because the competing methods use different loss functions for estimation. The prediction errors are therefore not directly comparable.

\section{Conclusion and discussion}\label{seccon}
We have developed a process that uses multiple loss functions separately to analyze linear models in the high dimensional data situation. Our approach has three notable advantages:
\begin{enumerate}[label=(\alph*)]
	\item it lowers the false discovery rate of variable selection by using our newly-developed vote procedure;
	
	\item it improves the quality of parameter estimation by combining results from multiple loss functions;
	
	\item it increases the speed of computation by adopting parallel computing.
	
\end{enumerate}
A specific instance of our approach, which optimally combines multiple quantile levels, achieves asymptotic efficiency of parameter estimation under mild conditions.

In practical applications some loss functions may be more relevant for model selection than others. Therefore a weighted version of the variable selector 
(\ref{vote}), namely
\bse
\hmq^{(w)}(\alpha)=\{j\in\{1,2,\ldots,p_n\}:\sk w_k I(\tvt_{kj}\neq 0)\geq\alpha\}
\ese
with a nonnegative weight vector $w=(w_1,\ldots,w_K)\trans$, may improve our vote procedure, which uses uniform weights $w_k=1$ for $k=1,\ldots,K$. Specifying and estimating the weight vector $w$ may be data-driven and vary from case to case. However, using uniform weights makes the vote procedure easier to implement. According to the results of Sections 4 and 5, this suffices to improve on competing methods with respect to variable selection.

\baselineskip=16pt

\section*{Supplementary materials}
\begin{itemize}
	\item All the programs in Sections \ref{secs} and \ref{secapp} are available at \url{https://github.com/guorongdai/Variable-Selection-through-Vote}.
	
	\item The data in Section \ref{secapp} are available at \url{https://www.ncbi.nlm.nih.gov/geo/geo2r/?acc=GSE5680}.
\end{itemize}

\section*{Acknowledgments}
The research of Dai and Carroll was supported by a grant from the National Cancer Institute (U01-CA057030).

\section*{Appendix}\label{app}

\noindent\underline{Proof of Theorem \ref{selection}}:
We will show the first conclusion of the theorem by proving that the two conditions in Lemma 1 of \citet{bradic2011penalized} hold true on an event with probability close to one. Let $\Psi_k(\theta)=\{\psi_k(Y_1-X_1\trans\theta),\ldots,\psi_k(Y_n-X_n\trans\theta)\}\trans$, $\gamma_k=(\gamma_{k1},\ldots,\gamma_{kp_n})\trans=X\trans\Psi_k(\vt)$ and
\be
\Gamma_{n,k}=\{\hbox{$\max_{j\in\mq^c}$}|\gamma_{kj}|\leq n^{1/2}z_n\}.
\label{Gamma}
\ee
Then we have
\bse
\p\{|\gamma_{kj}|>n^{1/2}z_n\}&=&\p\{|\si X_{ij}\psi_k(\varepsilon_i)|>n^{1/2}z_n\} \\
&\leq& 2\exp\{-(2n+2c\,n^{1/2}z_n)^{-1}nz_n^2\} \\
&=&2\exp[-\{2+2c\,n^{(\nu_0-2\nu_1)_+/2+\nu_2-1/2}\}^{-1}z_n^2] \\
&\leq& 2\exp(-c\,z_n^2).
\ese
In the above, the second step uses Lemma 2.2.11 of \citet{van1996weak} and the fact that $E\{\psi_k(\v_1)\}=0$ and $\max_{1\leq i\leq n}E\{|X_{ij}\psi_k(\varepsilon_i)|^m\}\leq  c\,m!T^{m-2}$ from Assumptions \ref{lfe} and \ref{dm2}. The last inequality uses Assumption \ref{tp}. It follows that
\be
\pr\{\Gamma_{n,k}\}\geq 1-\hbox{$\sum_{j\in\mq^c}$}\p\{|\gamma_{kj}|>n^{1/2}z_n\}\geq 1-2(p_n-q_n)\exp(-c\,z_n^2).
\label{p1}
\ee

Then, for $\tvt_k^o$ defined in (\ref{oracular}), with $d_{k\mq}=(d_{k1},\ldots,d_{kq_n})\trans$, we have
\be
\|\tvt_k^o-\vt\|&=&O_p(n^{-1/2}q_n^{1/2}+\lambda_{n,k}\|d_{k\mq}\|) \nonumber \\
&=&O_p(n^{-1/2}q_n^{1/2}+\lambda_{n,k}q^{1/2}D_{n,k}) \nonumber \\
&=&O_p(n^{-1/2}q_n^{1/2}+n^{-1/2})=O_p(n^{-1/2}q_n^{1/2})=O_p\{n^{(\nu_0-1)/2}\}. \label{distance}
\ee
Here the first step follows from Lemma 2 of \citet{bradic2011penalized}, the third step uses Assumption \ref{pw} and the last step uses Assumption \ref{dim}. The definition of $\tvt_k^o$ in (\ref{oracular}) implies that
\be
X_{\mq}\trans\Psi_k(\tvt^o_k)+n\lambda_{n,k}d_{k\mq}\circ\hbox{Sign}(\tvt^o_{k\mq})=\mathbf{0},
\label{nonzero}
\ee
where the bold number $\mathbf{0}$ denotes a $q_n$-dimensional vector whose components all equal zero, $\tvt^o_{k\mq}=(\tvt^o_{k1},\ldots,\tvt^o_{kq_n})\trans$, the symbol $\circ$ represents the Hadamard product and Sign$(\cdot)$ is taken componentwise. Here Sign$(x)=|x|^{-1}x$ for a scalar $x\neq 0$ and Sign$(0)\in[-1,1]$. With $\tilde{d}_{k\mq^c}=\{d_{k(q_n+1)}^{-1},\ldots,d_{kp_n}^{-1}\}\trans$ we have that on the event $\Gamma_{n,k}$ defined in (\ref{Gamma}),
\be
\|\tilde{d}_{k\mq^c}\circ X_{\mq^c}\trans\Psi_k(\tvt^o_k)\|_{\infty}&\leq &\|\tilde{d}_{k\mq^c}\circ X_{\mq^c}\trans\Psi_k(\vt)\|_{\infty}+\|\tilde{d}_{k\mq^c}\circ X_{\mq^c}\trans\{\Psi_k(\tvt^o_k)-\Psi_k(\vt)\}\|_{\infty} \nonumber \\
&\leq & c\,\{n^{1/2}z_n+\|X_{\mq^c}\trans G_k(\bvt_k)X_{\mq}(\tvt_k^o-\vt)\|_\infty\} \nonumber \\
&\leq& c\,\{n^{1/2}z_n+\|X_{\mq^c}\trans G_k(\bvt_k)X_{\mq}\|_{2,\infty}\|\tvt_k^o-\vt\|\} \nonumber \\
&\leq& c\,(n^{1/2}z_n+n^{1-\nu_1}\|\tvt_k^o-\vt\|) \nonumber \\
&=& O\{n^{(\nu_0-2\nu_1)_+/2+\nu_2+1/2}\}+O_p(n^{1-\nu_1})O_p\{n^{(\nu_0-1)/2}\} \nonumber \\
&=& O\{n^{(\nu_0-2\nu_1)_+/2+\nu_2+1/2}\}+O_p(n^{\nu_0/2-\nu_1+1/2})=o_p(n\lambda_{n,k}). \label{zero}
\ee
In the above, the second inequality uses (\ref{Gamma}), Assumption \ref{pw} and Taylor's expansion with $\bvt_k=\vt+\mu(\tvt_k^o-\vt)$ for some $\mu\in(0,1)$. The fourth step holds by Assumption \ref{dm1} and the fact that $\|\bvt_k-\vt\|<\|\tvt^o_k-\vt\|=O\{n^{(\nu_0-1)/2}\}$ from (\ref{distance}). The fifth step uses (\ref{distance}) and the last step follows from Assumption \ref{tp}.

The equations (\ref{nonzero}) and (\ref{zero}) gurantee the conditions (27) and (28) of Lemma 1 in \citet{bradic2011penalized} are satisfied, which implies that $\tvt_k^o$ is the unique global minimizer of the objective function in (\ref{ob}) on $\Gamma_{n,k}$. This combined with (\ref{p1}), the definition of $\tvt_k$ in (\ref{ob}) and Assumption \ref{tp}
implies
\be
\p\{\tvt_k=\tvt_k^o\}\geq\p\{\Gamma_{n,k}\}=1-2(p_n-q_n)\exp(-c\,z_n^2)\to 1.
\label{match}
\ee
This gives the first conclusion of the theorem. The equation (\ref{match}) and the definition of $\tvt_k^o$ in (\ref{oracular}) further yield
\be
\p\{\hbox{$\cap_{j\in\mq^c}$}\{\tvt_{kj}=\vt_j=0\}\}\to 1.
\label{prezero}
\ee
Moreover, we know that, with probability tending to one,
\be
\|\tvt_{k\mq}-\vt_{\mq}\|=\|\tvt_k-\vt\|=\|\tvt^o_k-\vt\|=O_p\{n^{(\nu_0-1)/2}\}.
\label{neg}
\ee
In the above the first two steps use (\ref{prezero}) and (\ref{match}) and the last step follows from (\ref{distance}). Then we have
\be
\p\{\hbox{$\cap_{j\in\mq}$}\{|\tvt_{kj}|>0\}\}&\geq& \p\{\hbox{$\cap_{j\in\mq}$}\{|\tvt_{kj}|> |\vt_{j}|-\hbox{$\min_{j\in\mq}$}|\vt_j|\}\}  \nonumber \\
&\geq & \p\{\hbox{$\cap_{j\in\mq}$}\{|\tvt_{kj}-\vt_j|< \hbox{$\min_{j\in\mq}$}|\vt_j|\} \} \nonumber \\
&\geq&\p\{\|\tvt_{k\mq}-\vt_{\mq}\|< \hbox{$\min_{j\in\mq}$}|\vt_j|\}\to 1, \label{prenonzero}
\ee
with $\tvt_{k\mq}=(\tvt_{k1},\ldots,\tvt_{kq_n})\trans$, where the convergence follows from (\ref{neg}) and Assumption \ref{sig}.

 Combining (\ref{prezero}) and (\ref{prenonzero}) yields $\p\{\hat{\mathcal{Q}}_{n,k}=\mq\}\to 1$ for $k=1,\ldots,K$, where $\hat{\mathcal{Q}}_{n,k}=\{j\in\{1,2,\ldots,p_n\}:\tvt_{kj}\neq 0\}$. It follows that
 \bse
 \hbox{$\pr\{\cap_{j\in\mq}\{\sk I(\tvt_{kj}\neq 0)=K\}\}\to 1$ and $\pr\{\cap_{j\in\mq^c}\{\sk I(\tvt_{kj}\neq 0)=0\}\}\to 1$.}
 \ese
By the definition of $\hmq(\alpha)$ in (\ref{vote}), we have $\p\{\hmq(\alpha)=\mq\}\to 1$ for any positive integer $\alpha\leq K$.

\noindent\underline{Proof of Theorem \ref{normality}}:
For $k=1,\ldots,K$ set
\bse
\hvt^o_{k\mq}=\hbox{$\mathop{\arg\min}_{\theta}$}\si\ell_k(Y_i-X_{i\mq}\trans\theta).
\ese
From Theorem \ref{selection} we have $\pr\{\hvt_{k\mq}=\hvt_{k\mq}^o\}\to 1$ for $k=1,\ldots,K$, which implies that
\be
\pr\{\hvt_{\mq}(w)=\sk w_k\hvt^o_{k\mq}\}\to 1.
\label{equiv}
\ee

For $k=1,\ldots,K$, Theorem 2.2 and Example 1 of \citet{he2000parameters} give that
\bse
\hvt^o_{k\mq}-\vt_{\mq}=-(\eta_k X_{\mq}\trans X_{\mq})^{-1}\si\psi_k(\varepsilon_i)X_{i\mq}+\varpi_n
\ese
with $\|\varpi_n\|=o_p(n^{-1/2})$. It follows that
\be
\label{expansion}
v\trans(X_{\mq}\trans X_{\mq})^{1/2}(\sk w_k\hvt^o_{k\mq}-\vt_{\mq})=\si L_{n,i}+o_p(1)
\ee
with $L_{n,i}=-v\trans( X_{\mq}\trans X_{\mq})^{-1/2}X_{i\mq}\sk w_k \eta_k^{-1} \psi_k(\varepsilon_i)$, which holds because
\bse
|v\trans(X_{\mq}\trans X_{\mq})^{1/2}\varpi_n|\leq n^{1/2}\lambda_{\max}^{1/2}(n^{-1}X_{\mq}\trans X_{\mq})\|v\|\,\|\varpi_n\|\leq c\,n^{1/2}\|\varpi_n\|=o_p(1).
\ese
The second step in the above uses Assumption \ref{dm2}. For $i=1,\ldots,n$, we have
\be
\label{mean}
E(L_{n,i})=0,
\ee
since $E\{\psi_k(\varepsilon_i)\}=0$ from Assumption \ref{lfe}. Then we compute
\be
\si E(L_{n,i}^2)&=&E[\{\sk w_k \eta_k^{-1} \psi_k(\varepsilon_1)\}^2]\si \{v\trans( X_{\mq}\trans X_{\mq})^{-1/2}X_{i\mq}\}^2 \nonumber \\
&=& (w\trans Hw)v\trans( X_{\mq}\trans X_{\mq})^{-1/2}\{\si(X_{i\mq}X_{i\mq}\trans)\}( X_{\mq}\trans X_{\mq})^{-1/2}v \nonumber \\
&=&w\trans Hw. \label{variance}
\ee
We have, for any $\zeta>0$,
\be
\si E\{L_{n,i}^2I(|L_{n,i}|>\zeta)\}&\leq& c\,\si E(L^4_{n,i}) \nonumber \\
&=&c\,E[\{\sk w_k \eta_k^{-1} \psi_k(\varepsilon_1)\}^4] \si \{v\trans( X_{\mq}\trans X_{\mq})^{-1/2}X_{i\mq}\}^4\nonumber \\
&\leq&c\,\si \{X_{i\mq}\trans( X_{\mq}\trans X_{\mq})^{-1/2} vv\trans ( X_{\mq}\trans X_{\mq})^{-1/2} X_{i\mq}\}^2 \nonumber \\
&\leq&c\,\lambda^2_{\max}(vv\trans)\si \{X_{i\mq}\trans( X_{\mq}\trans X_{\mq})^{-1}  X_{i\mq}\}^2 \nonumber \\
&=&c\,\si \{X_{i\mq}\trans( X_{\mq}\trans X_{\mq})^{-1} X_{i\mq}\}^2 \nonumber \\
&\leq&c\,n^{-2}\lambda_{\min}^{-2}(n^{-1}X\trans_{\mq}X_{\mq})\si \|X_{i\mq}\|^4 \nonumber \\
&\leq&c\,n^{-1} q_n^2\|X_{\mq}\|^4_\infty\leq c\,n^{-1}q_n^2=o(1). \label{lf}
\ee
Here the third step uses the facts that $E\{\psi_k(\varepsilon_1)^4\}\leq c$ (Assumption \ref{lfe}) and that $\eta_k> 0$ (Assumption \ref{lfe}), the fifth step holds true because $\|v\|=1$, the seventh and the eighth steps use Assumption \ref{dm2}, and the last step uses the fact that $q_n=o(n^{1/2})$.  This shows that the Lindeberg-Feller condition for the central limit theorem is satisfied. Summing up, the equations (\ref{expansion}) through (\ref{lf}) yield
\bse
v\trans(X_{\mq}\trans X_{\mq})^{1/2}(\sk w_k\hvt^o_{k\mq}-\vt_{\mq})\xrightarrow{d}N(0,w\trans H w).
\ese
This combined with (\ref{equiv}) completes the proof.

\noindent\underline{Proof of Corollary \ref{oweight}}: Since $w^*=\arg\min_{w:\,r\trans w=1}(w\trans Hw)$ we have, by the Lagrange multiplier method, that $w^*=(r\trans H^{-1}r)^{-1}(H^{-1}r)$. Theorem \ref{normality}, with $w=w^*$ in the asymptotic variance formula, gives
\be
v\trans (\xq\trans \xq)^{1/2}\{\hvt_{\mq}(w^*)-\vt_{\mq}\}\stackrel{d}{\longrightarrow}N\{0,(r\trans H^{-1}r)^{-1}\}. \label{optdistribution}
\ee

Let $\home$ be a consistent estimator of $w^*$. Then
\be
&&\phantom{=}|v\trans (\xq\trans \xq)^{1/2}\{\hvt_{\mq}(\home)-\vt_{\mq}\}-v\trans (\xq\trans \xq)^{1/2}\{\hvt_{\mq}(w^*)-\vt_{\mq}\}| \nonumber \\
&&=|\sk(w^*_k-\home_k)v\trans(X_{\mq}\trans X_{\mq})^{1/2}(\hvt_{k\mq}-\vt_{\mq})| \nonumber \\
&&\leq (\hbox{$\max_{1\leq k\leq K}$}|w^*_k-\home_k|)\sk|v\trans(X_{\mq}\trans X_{\mq})^{1/2}(\hvt_{k\mq}-\vt_{\mq})|=o_p(1). \label{diff}
\ee
The last step uses the fact that $|v\trans(X\trans X)^{1/2}(\hvt_{k\mq}-\vt_{\mq})|=O_p(1)$ for $1\leq k\leq K$, which holds by Theorem \ref{normality}, and the consistency of $w^*$.
Combining (\ref{optdistribution}) and (\ref{diff}) yields
\bse
v\trans (\xq\trans \xq)^{1/2}\{\hvt_{\mq}(\home)-\vt_{\mq}\}\stackrel{d}{\longrightarrow}N\{0,(r\trans H^{-1}r)^{-1}\}.
\ese

\noindent\underline{Proof of Theorem \ref{efficiency}}: The $i\th$ diagonal element of the $K \times K$ matrix $H^{-1}$ is $2(K+1)\{f(\beta_i)\}^2$ for $i=1,\dots,K$. The $(i,i+1)\th$ and $(i+1,i)\th$ entries are $-(K+1)f(\beta_i)f(\beta_{i+1})$ for $i=1,\dots,K-1$, and the other entries are zero. Hence we have
\be
r\trans H^{-1}r&=&2(K+1)[\sk \{f(\beta_k)\}^2-\hbox{$\sum_{k=1}^{K-1}$}f(\beta_k)f(\beta_{k+1})] \nonumber\\
&=&2(K+1)[\sk \{g(\tau_k)\}^2-\hbox{$\sum_{k=1}^{K-1}$}g(\tau_k)g(\tau_{k+1})] \nonumber\\
&=&(K+1)[\hbox{$\sum_{k=1}^{K-1}$}\{g(\tau_{k+1})-g(\tau_{k})\}^2+\{g(\tau_1)\}^2+\{g(\tau_K)\}^2] \nonumber\\
&=&(K+1)[\{g(\tau_1)\}^2+\{g(\tau_K)\}^2]+\hbox{$\int_{\tau_1}^{\tau_K}$}\{g'(t)\}^2dt+L_K, \label{decomp}
\ee
where $L_K=(K+1)\sum_{k=1}^{K-1}\{g(\tau_{k+1})-g(\tau_{k})\}^2-\int_{\tau_1}^{\tau_K}\{g'(t)\}^2dt$ with
\be
|L_K|&=&|(K+1)\hbox{$\sum_{k=1}^{K-1}$}[\{\hbox{$\int_{\tau_k}^{\tau_{k+1}}$}g'(t)dt\}^2-(\tau_{k+1}-\tau_k)\hbox{$\int^{\tau_{k+1}}_{\tau_{k}}$}\{g'(t)\}^2dt]| \nonumber \\
&=&\{(K+1)/2\}\hbox{$\sum_{k=1}^{K-1}\int^{\tau_{k+1}}_{\tau_{k}}\int^{\tau_{k+1}}_{\tau_{k}}\{g'(x)-g'(y)\}^2dxdy$}\nonumber \\
&\leq&\{(K+1)/2\}\hbox{$\sum_{k=1}^{K-1}$}(\tau_{k+1}-\tau_k)^2\hbox{$\max_{x,y\in[\tau_{k},\tau_{k+1}]}$}\{g'(x)-g'(y)\}^2\nonumber \\
&=&\{2(K+1)\}^{-1}\hbox{$\sum_{k=1}^{K-1}$}\hbox{$\max_{x,y\in[\tau_{k},\tau_{k+1}]}\{\int^x_y$}g''(t)dt\}^2\nonumber \\
&\leq&\{2(K+1)\}^{-1}\hbox{$\sum_{k=1}^{K-1}$}\hbox{$\{\int^{\tau_{k+1}}_{\tau_k}$}|g''(t)|dt\}^2\nonumber \\
&\leq&\{2(K+1)^2\}^{-1}\hbox{$\sum_{k=1}^{K-1}$}[\hbox{$\int^{\tau_{k+1}}_{\tau_k}$}\{g''(t)\}^2dt]\nonumber \\
&=&\{2(K+1)^2\}^{-1}\hbox{$\int^{\tau_{K}}_{\tau_1}$}\{g''(t)\}^2dt\to 0 \quad (K \to \infty). \label{vanish}
\ee
The sixth step in the above uses the Cauchy-Schwarz inequality and the last step follows from Assumption \ref{density}. Then, using (\ref{decomp}), we obtain
\bse
&&\phantom{=}|r\trans H^{-1}r-\hbox{$\int_{\mf}$}\{f(t)\}^{-1}\{f'(t)\}^2dt-\nu_3| \\
&&=|(K+1)[\{g(\tau_1)\}^2+\{g(\tau_K)\}^2]+\hbox{$\int_{\tau_1}^{\tau_K}$}\{g'(t)\}^2dt+L_K-\hbox{$\int^1_{0}$}\{g'(t)\}^{2}dt-\nu_3| \\
&&\leq|(K+1)[\{g(\tau_1)\}^2+\{g(\tau_K)\}^2]-\nu_3|+|L_K| +\hbox{$\int^{\tau_{1}}_{0}$}[\{g'(t)\}^2+\{g'(1-t)\}^2]dt \\
&&\to 0 \quad \ (K \to \infty).
\ese
The last step follows from Assumption \ref{density}, the equation (\ref{vanish}) and the fact that $\tau_1=(K+1)^{-1}\to 0$ as $K\to\infty$. This completes the proof.
\bibliographystyle{biometrika}
\bibliography{myreference-vsv}
\end{document}